%% file: GNdraft.tex
\documentclass[%
 reprint,
superscriptaddress,
 amsmath,amssymb,
 aps,
prx,
]{revtex4-2}

\usepackage{graphicx}
\usepackage{dcolumn}
\usepackage{siunitx}
 \usepackage{relsize}
\usepackage{placeins}
\usepackage{subfig}
\usepackage{caption}
\usepackage{newtxtext,newtxmath}
\usepackage{xcolor} 
\usepackage{float}
\usepackage{isotope}
\usepackage{hyperref}
\hypersetup{
    colorlinks=true,
    linkcolor=blue,
   urlcolor=blue,
citecolor=blue}

\DeclareMathAlphabet{\mathcal}{OMS}{cmsy}{m}{n}

\makeatletter
\let\save@mathaccent\mathaccent
\newcommand*\if@single[3]{%
  \setbox0\hbox{${\mathaccent"0362{#1}}^H$}%
  \setbox2\hbox{${\mathaccent"0362{\kern0pt#1}}^H$}%
  \ifdim\ht0=\ht2 #3\else #2\fi
  }
\newcommand*\rel@kern[1]{\kern#1\dimexpr\macc@kerna}
\newcommand*\widebar[1]{\@ifnextchar^{{\wide@bar{#1}{0}}}{\wide@bar{#1}{1}}}
\newcommand*\wide@bar[2]{\if@single{#1}{\wide@bar@{#1}{#2}{1}}{\wide@bar@{#1}{#2}{2}}}
\newcommand*\wide@bar@[3]{%
  \begingroup
  \def\mathaccent##1##2{%
    \let\mathaccent\save@mathaccent
    \if#32 \let\macc@nucleus\first@char \fi
    \setbox\z@\hbox{$\macc@style{\macc@nucleus}_{}$}%
    \setbox\tw@\hbox{$\macc@style{\macc@nucleus}{}_{}$}%
    \dimen@\wd\tw@
    \advance\dimen@-\wd\z@
    \divide\dimen@ 3
    \@tempdima\wd\tw@
    \advance\@tempdima-\scriptspace
    \divide\@tempdima 10
    \advance\dimen@-\@tempdima
    \ifdim\dimen@>\z@ \dimen@0pt\fi
    \rel@kern{0.6}\kern-\dimen@
    \if#31
      \overline{\rel@kern{-0.6}\kern\dimen@\macc@nucleus\rel@kern{0.4}\kern\dimen@}%
      \advance\dimen@0.4\dimexpr\macc@kerna
      \let\final@kern#2%
      \ifdim\dimen@<\z@ \let\final@kern1\fi
      \if\final@kern1 \kern-\dimen@\fi
    \else
      \overline{\rel@kern{-0.6}\kern\dimen@#1}%
    \fi
  }%
  \macc@depth\@ne
  \let\math@bgroup\@empty \let\math@egroup\macc@set@skewchar
  \mathsurround\z@ \frozen@everymath{\mathgroup\macc@group\relax}%
  \macc@set@skewchar\relax
  \let\mathaccentV\macc@nested@a
  \if#31
    \macc@nested@a\relax111{#1}%
  \else
    \def\gobble@till@marker##1\endmarker{}%
    \futurelet\first@char\gobble@till@marker#1\endmarker
    \ifcat\noexpand\first@char A\else
      \def\first@char{}%
    \fi
    \macc@nested@a\relax111{\first@char}%
  \fi
  \endgroup
}
\makeatother

\newsavebox{\mybox}
\newlength{\mywidth}
\newlength{\myheight}
\newlength{\myline}
\newlength{\myoffset}
\newcommand{\mysqrt}[1]%
{\setlength{\myline}{.1ex}%
\addtolength{\myline}{.06pt}%
\setlength{\myoffset}{.9em}
\addtolength{\myoffset}{-2pt}
\savebox{\mybox}{$\displaystyle\sqrt{#1}$}%
\settoheight{\myheight}{\usebox{\mybox}}%
\addtolength{\myheight}{-.2ex}
\settowidth{\mywidth}{\usebox{\mybox}}%
\addtolength{\mywidth}{-\myoffset}%
 \rlap{\usebox{\mybox}}\hspace{\myoffset}{\raisebox{\myheight}{\rule{\mywidth}{\myline}}}}

\begin{document}

\preprint{APS/123-QED}

\title{Limitations of the Geiger-Nuttall law in heavy cluster decay}

\author{Omar Nagib}%
\email{{\color{black} omar.nagib@aucegypt.edu}}
\affiliation{Department of Physics, American University in Cairo, Cairo 11835, Egypt}
\author{Ahmed M. Hamed}%
\affiliation{Department of Physics, American University in Cairo, Cairo 11835, Egypt}
\affiliation{Department of Physics and Astronomy, Texas A\&M University, College Station, TX 77843, USA}
\affiliation{Department of Physics and Astronomy, University of Mississippi, Oxford, MS 38677, USA}

\date{\today}

\begin{abstract}
Geiger-Nuttall law is the simplest relation in radioactive decay relating the half-life and the decay energy. Initially restricted to $\alpha$ decay of individual isotopes, generalizations unifying different isotopes and decay modes under a single set of constant coefficients were subsequently achieved. This motivates investigating to what extent such generalizations are possible. We also examine whether there exists a universal Geiger-Nuttall law that can simultaneously describe all decay modes and nuclei including heavy clusters. We show that the validity of Geiger-Nuttall law and its generalizations hinges on the assumption that half-life can be approximated linearly as a function of the square root of the ratio of the decay energy to the Coulomb barrier height. Systematic calculation of the ratio across the nuclear chart for 12 decay modes reveals that it varies over its whole range between 0 and 1. Consequently, no linear approximation can unify all the nuclei and decay modes under a single set of coefficients, and thus no universal Geiger-Nuttall law is possible in contrast to previous claims. In cluster decay, the ratio varies within 0.6--1 where non-linearity becomes significant such that no generalized Geiger-Nuttall description of heavy clusters is possible. In the ongoing attempts of unification, it might be necessary to go beyond the Geiger-Nuttall law and incorporate additional terms proportional to the decay energy and/or its square root.

\end{abstract}

\maketitle


\section{\label{intro}Introduction}

Ever since the inception of nuclear physics, the study of charged-particle radioactivity was always in the field's forefront. Historically, this started with the discovery of $\alpha$ decay \cite{rutherford}, followed by proton emission \cite{proton1, proton2}, and nuclei heavier than helium which is known as cluster decay. So far, the clusters $\isotope[14]{C}, \isotope[20]{O}, \isotope[23]{F}, \isotope[22,24-26]{Ne},\isotope[28,30]{Mg}$ and $\isotope[32,34]{Si}$ have been observed to decay from parent nuclei between $\isotope[221]{Fr}$ and $\isotope[242]{Cm}$ \cite{carbon,24Ne,NeMg,14CRa, FeNe,O,34Si,22Ne}. The great interest in charged-particle radioactivity is justified since it provides myriads of invaluable information about the nuclear structure that is otherwise difficult to obtain through a detailed microscopic analysis of every nucleus. The half-life of the nucleus and the decay energy is the experimentally measurable quantities in radioactive decay. Remarkably enough, just with the aid of these two quantities, it is possible to learn about nuclear stability, proton, and neutron shell closures, nuclei deformation, nuclear charge radii, and nucleons clusterization inside the nuclei \cite{shell,GNp3, deform2, radius, preform1, preform2}. This is the primary motivation historically and currently behind the development of theoretical models that can reproduce and predict decay half-lives accurately and systematically across the nuclear chart.  

In particular, Geiger-Nuttall law proves to be one of the most versatile tools among the wealth of models available for half-life systematics. The law relates the $\alpha$-decay half-life $T_{1/2}$ to the $\alpha$-decay energy $Q_{\alpha}$ by a simple relation:

\begin{equation} \label{GN}
\log_{10}T_{1/2}=a Q_{\alpha}^{-1/2}+b
\end{equation}

where $a$ and $b$ are constants. Both the physical mechanism of $\alpha$ decay and the $Q_{\alpha}^{-1/2}$ dependence were explained as a consequence of quantum tunneling of the $\alpha$ particle trapped inside the parent nuclei through the Coulomb barrier in landmark papers by Gamow \cite{gamow} and Gurney and Condon \cite{condon1, condon2}. While it was discovered more than 100 years ago \cite{GN}, it remains of general validity for all known nuclei reproducing experimental half-life within one order of magnitude or less with few exceptions (e.g., $\isotope[186]{Po}$) \cite{GNvalid}. It turns out, however, that the coefficients $a$ and $b$ change for every isotopic chain and change for a given isotopic chain when magic numbers are crossed (e.g., $N=126$) \cite{GNmagic}. Fifty years later, Viola and Seaborg successfully extended the law for all isotopic chains by incorporating a dependence on the parent proton number $Z$ \cite{VS}. Plenty of analogous formulas were proposed all sharing the characteristic $Q_{\alpha}^{-1/2}$ dependence \cite{royer,SP, SP2, denv, santosh, brown, TQstudy}. Geiger-Nuttall law was subsequently found to hold for proton \cite{GNp1,GNp2,GNp3,GNp4} and cluster radioactivity as well \cite{GNc1, GNc2, GNc3, horoi} which is expected since they share the same tunneling mechanism with $\alpha$ decay. 

The last two decades have witnessed great progress on the development of generalizations of Geiger-Nuttall law for various decay modes accompanied by an increased theoretical understanding of these laws. Arguably, the biggest breakthrough was the unification of $\alpha$ and all experimentally known cluster decay modes (i.e., from \isotope[14]{C} up to $\isotope[34]{Si}$) with a single generalized Geiger-Nuttall law. The universal decay law (UDL) \cite{UDL1, UDL2}, and the unified formula (UF) \cite{UF} are two such examples describing $\alpha$ and cluster decay of all isotopes with a single set of fitted coefficients. Moreover, they can successfully reproduce proton emission half-lives when additional angular momentum and deformation terms are taken into account \cite{GNp3, UFp}. The significance of these formulas is that they were derived from general theoretical frameworks (i.e., the R-matrix theory and the square-well model) that are valid for all decay modes and nuclei. This led the authors of the UDL to claim that it should be valid for all decay modes and isotopic chains \cite{UDL1, UDL2, GNreview, GNreview2} with a similar claim of unification made by the authors of the UF \cite{UF, UFp}. The series of historically successful unifications encompassing an ever-larger number of decay modes and isotopic chains prompts the question: Is there a universal Geiger-Nuttall law encompassing all nuclei and decay modes including clusters much heavier than $\isotope[34]{Si}$?

Recent calculations from various models (e.g., fission, microscopic, and liquid drop models) suggest that radioactive decay of clusters much heavier than $\isotope[34]{Si}$ may become a dominant mode in superheavy nuclei with $Z > 110$ alongside $\alpha$ decay \cite{HCR1, HCR2, HCR3, HCR4}. On the other hand, systematic studies of $\alpha$ and cluster decay using different generalized Geiger-Nuttall laws (e.g., UDL, UF, and Horoi formula) reveal extremely large discrepancies in their predictions and contradictory conclusions about the behavior of half-life and branching ratios in heavy cluster decay \cite{GNpredict1, GNpredict2, GNpredict3}. Thus it is also of practical interest to investigate the validity of generalized Geiger-Nuttall laws in describing and predicting heavy cluster decay and explain these anomalies. More generally, this paper constitutes a study of the assumptions required to ensure the validity of the three classes of Geiger-Nuttall laws: 

\begin{enumerate} 

\item Simple Geiger-Nuttall law [Eq. \eqref{GN}] which describes a single isotopic chain and a decay mode. 
\item Generalized Geiger-Nuttall laws which describe more than a single isotopic chain and/or decay mode (e.g., Viola-Seaborg law, UDL, and UF). 
\item Universal Geiger-Nuttall law which describes all decay modes and nuclei simultaneously with a single set of coefficients.
\end{enumerate} 

This paper is organized as follows. In Sec. \ref{model}, we use the R-matrix theory to show the assumptions required to arrive at Geiger-Nuttall laws. The most crucial assumption---dubbed the linearity assumption---is identified, namely, that $x$ varies within an interval such that half-life can be approximated linearly in $x$ which is the square root of the ratio of the decay energy $Q$ to the Coulomb barrier height. In Sec. \ref{results}, the variation of $x$ with nuclei and decay modes and its implication for the validity of the three classes of Geiger-Nuttall laws is discussed. $x$ of 5130 heavy and superheavy nuclei with 12 cluster decay modes from the proton up to $\isotope[54]{Ti}$ are computed using the experimental $Q$ and ones calculated by the WS4 mass model with the radial basis function correction. It is found that $x$ varies over its whole possible range between 0 and 1 across nuclei and decay modes and thus a universal Geiger-Nuttall law is not possible. The success of the simple Geiger-Nuttall law in $\alpha$ decay and its generalizations (e.g., Viola-Seaborg law) are explained owing to $x$ being always smaller than 0.6 where the linearity assumption is true. Next, the discrepancies arising between different generalized Geiger-Nuttall laws in heavy cluster decay is explained using the insights gained about the $x$ variation. Finally, it is found that any generalized Geiger-Nuttall description of heavy clusters is not valid since the role of non-linearity becomes significant. Section \ref{conclusions} concludes the paper. 

\section{Theoretical framework} \label{model}

Here we build on the derivations from Refs. \cite{UDL1, UDL2} to scrutinize and test the various assumptions underlying Geiger-Nuttall laws. The R-matrix theory provides a general microscopic framework valid for all cluster radioactivity and nuclei where decay is understood as a two-step mechanism beginning with the formation of the cluster inside the parent nuclei followed by tunneling through the Coulomb barrier. Thomas derived an expression for the decay width $\Gamma$ valid for the proton, $\alpha$, cluster radioactivity and all nuclei given by \cite{Rmatrix1}

\begin{equation} \label{width}
\Gamma=2 P_l \delta^2_l
\end{equation} 

where $P_l$ is the penetrability (i.e., tunneling probability) given by

\begin{equation} \label{Pl}
P_l=\dfrac{kR}{|H_{l}^{+}(\eta,kR)|^2}
\end{equation}

$H_{l}^{+}(\eta,kR)$ is the Coulomb-Hankel function at distance $R$ between the cluster and the daughter nuclei parametrized by the angular momentum $l$ and the Coulomb parameter $\eta$:

\begin{equation} \label{eta}
\eta=\dfrac{2 e^2 Z_c Z_d}{\hbar v}
\end{equation}

where $Z_c$ and $Z_d$ are the proton numbers of the cluster and the daughter, respectively. $k=\mu v/\hbar$ where $v$ is the velocity of the emitted cluster related to the decay energy $Q$ and the reduced mass $\mu$ of the cluster-daughter system by $Q= \mu v^2/2$. $\eta$ can be alternatively described in terms of $\mu$ and $Q$ as

\begin{equation} \label{eta2}
\eta=\dfrac{ e^2 Z_c Z_d}{\hbar} \sqrt{\dfrac{2 \mu}{Q}}
\end{equation}

 $\delta^2_l$ is the reduced width describing the formation of the clusters inside the parent nuclei given by 

\begin{equation} \label{rwidth}
\delta_l= \sqrt{\dfrac{\hbar^2 R}{2 \mu}}g_l(R)
\end{equation}

where $g_l(R)$ is the cluster formation amplitude at a distance $R$. Combining Eqs. \eqref{width}, \eqref{Pl}, and \eqref{rwidth}, $T_{1/2}$ will be given by 

\begin{equation} \label{T}
T_{1/2}=\dfrac{\hbar \ln2}{\Gamma}=\dfrac{\ln 2}{v} \bigg| \dfrac{H_{l}^{+}(\eta,kR) }{R g_l(R)} \bigg|^2
\end{equation} 

While both $H_{l}^{+}(\eta,kR)$ and $Rg_l(R)$ depend on $R$, their ratio, and therefore, $T_{1/2}$ (and $\Gamma$) are independent of $R$ for large distances where nuclear interaction is negligible \cite{UDL1, UDL2, GNreview, nucdecay}. Following previous authors, we take $R=r_0(A^{1/3}_c+A^{1/3}_d)$ as the touching radius where $A_c$ and $A_d$ are the mass numbers of the cluster and the daughter nuclei, respectively \cite{GNp1, UDL1, UDL2, Gmodel, Gmodelp,UNIV}. We take $r_0 \approx 1.2 \ \si{\femto \meter}$ since it can best reproduce the experimental nuclear charge radii as well as the half-life data of all decay modes \cite{ r0, r02, Gmodel, Gmodelp}. $R$ should not be taken smaller than the touching radius since the nuclear interaction starts to become significant when there is an overlap between the cluster and the daughter nuclei. To arrive at the simple Geiger-Nuttall law [Eq. \eqref{GN}] or generalizations thereof from the general Eq. \eqref{T} above, it is required to obtain an analytical expression for $H_{l}^{+}$ and $Rg_l(R)$. $|H_{l}^{+}(\eta,kR)|^2$ is given by \cite{GNp1, Hderive} 

\begin{equation} \label{CH}
|H_{l}^{+}(\eta, k R)|^2= \zeta_l(x) \exp[2 \gamma_l(x)]
\end{equation} 

where $\gamma_l(x)$ is given by

\begin{equation} \label{gamma}
\gamma_l(x)= \eta \big( \arccos(x) -x \sqrt{1-x^2} \big ) +\dfrac{l(l+1)}{\eta} \dfrac{\sqrt{1-x^2}}{x}
\end{equation} 

and  $\zeta_l(x)$ is given by

\begin{equation} \label{zeta}
\zeta_l(x)= \bigg( \dfrac{1-x^2}{x^2} +\dfrac{l(l+1)}{(kR)^2} \bigg)^{-1/2}
\end{equation} 

where $0 < x < 1$ is the dimensionless ratio given by 

\begin{equation} \label{x}
x= \sqrt{\dfrac{kR}{\eta}}=\sqrt{ \dfrac{Q}{V_C(R)} }=\sqrt{\dfrac{QR}{e^2 Z_c Z_d}}=\sqrt{\dfrac{R}{R_C}}
\end{equation} 

where $V_C(R)=e^2 Z_c Z_d/R$ and $R_C= e^2 Z_c Z_d/Q$ is the classical turning point known as the Coulomb radius. An intuitive physical picture of $x$ is the square root of the ratio of the first classical turning point $R$ to the second $R_C$ in the square-well model the width of which is $R$. Equivalently, it is the square root of the ratio of $Q$ to the Coulomb barrier height $V_C(R)$. $x$ is not defined for $Q<0$ (i.e., nuclei stable against decay) since the second classical turning point $R_C$ satisfying $V_C(R_C)=Q<0$ does not exist. Taking the logarithm of $T_{1/2}$ [Eq. \eqref{T}] and substituting the expression for $|H_{l}^{+}(\eta, k R)|^2$ [Eq. \eqref{CH}] we get

\begin{equation} \label{logT}
\log_{10}T_{1/2}=  \frac{2}{\ln 10} \gamma_l(x) + d - \log_{10} |R g_l(R)|^2
\end{equation}

where the second term $d=\log_{10}  \zeta_l(x) \ln2 /v $ varies smoothly for all decay modes and nuclei such that it can be considered a constant (see Appendix \ref{assumptions} for discussion). 

The most crucial assumption that leads to any form of Geiger-Nuttall law, simple or generalized, relates to $\gamma_l(x)$, namely, that it can be approximated as a linear function of $x$. This assumption is necessary so that $\log_{10}{T_{1/2}}$ gives rise to the characteristic $Z_c Z_d Q^{-1/2}$ dependence without any other $Q$ dependencies due to higher-order terms in $x$. Thus for the sake of the argument, assume that $x$ varies around a certain point $x=x_0$ in a sufficiently small interval such that without loss of accuracy, we can expand the $\arccos (x)- x\sqrt{1-x^2}$ term in $\gamma_l(x)$ [Eq. \eqref{gamma}] in a Taylor series at $x=x_0$ retaining only terms up to first order in $x$:

\begin{multline} \label{taylor}
\gamma_l(x)= \eta \big( \arccos(x) -x \sqrt{1-x^2} \big ) +\dfrac{ l(l+1)}{\eta} \dfrac{\sqrt{1-x^2}}{x} \\ \approx  \eta (a_0 +b_0 x) +\dfrac{\hbar l(l+1)}{\sqrt{2 e^2 \mu R Z_c Z_d}}\sqrt{1-x_0^2}
\end{multline}

where $a_0$ and $b_0$ are the constant coefficients of the Taylor expansion at $x=x_0$ given by

\begin{equation} \label{a}
a_0= \arccos (x_0)+ x_0 \sqrt{1-x_0^2}
\end{equation}

\begin{equation} \label{b}
b_0= -2 \sqrt{1-x_0^2}
\end{equation}

In the denominator of the $l(l+1)$ term, we have used the fact that the product of $\eta$ [Eq. \eqref{eta2}] and $x$ [Eq. \eqref{x}] is independent of $Q$ and given by

\begin{equation} \label{etax}
\eta x= \dfrac{\sqrt{2 e^2 \mu R Z_c Z_d}}{\hbar}
\end{equation}

Meanwhile, in the numerator of the $l(l+1)$ term, we only retained terms up to zeroth order in $x$ for $\sqrt{1-x^2}$ since higher-order terms will give higher-order $Q$ dependencies different from $Q^{-1/2}$ of Geiger-Nuttall law. The first term in $ \eta (a_0 +b_0 x)$ [Eq. \eqref{taylor}] will give the $Q^{-1/2}$ dependence while the second term is independent of $Q$ since it involves the product of $\eta$ and $x$. Thus using Eqs. \eqref{eta2} and \eqref{etax} in $\eta (a_0 +b_0 x)$ we get

\begin{equation} \label{gamma0}
\eta (a_0 +b_0 x)= \dfrac{ e^2 a_0 }{\hbar }Z_c Z_d \sqrt{\dfrac{2 \mu}{Q}} + \dfrac{  e b_0}{\hbar } \sqrt{ 2 \mu R Z_c Z_d}
\end{equation}

By substituting Eqs. \eqref{gamma0} and \eqref{taylor} in Eq. \eqref{logT} we finally obtain

\begin{widetext}

\begin{equation} \label{UGN}
\log_{10}T_{1/2}=   a Z_c Z_d \sqrt{\dfrac{\mathcal{A}}{Q}} + b \sqrt{ \mathcal{A} (A_c^{1/3}+A_d^{1/3}) Z_c Z_d } +c \dfrac{l(l+1)}{ \sqrt{ \mathcal{A} (A_c^{1/3}+A_d^{1/3}) Z_c Z_d }} + d - \log_{10} |R g_l(R)|^2
\end{equation}

The substitutions $\mu=m \mathcal{A}$ where $m$ is the nucleon mass, $\mathcal{A}=A_c A_d/(A_c+A_d)$, and $R=r_0 (A_c^{1/3}+A_d^{1/3})$ were made above and all the constants have been lumped into the constant coefficients $a,b$, and $c$. Another implication of the linearity assumption of $\gamma_l(x)$ is that the formation probability $\log_{10} |R g_l(R)|^2$ should be proportional to $\sqrt{ \mathcal{A} (A_c^{1/3}+A_d^{1/3}) Z_c Z_d }$ as argued by the authors of the universal decay law \cite{UDL1, UDL2}. The argument relied on two assumptions, namely 1) the ratio $H_{l}^{+}(\eta,kR) / Rg_l(R)$ is independent of $R$ and 2) higher-orders terms in the expansion of $\gamma_l(x)$ (e.g., the $x^3$ term) vary smoothly across different nuclei and decay modes such that they can be considered constant (i.e., linearity assumption). Therefore, the above equation becomes (we will examine this implication later) 

\begin{equation} \label{UGN2}
\log_{10}T_{1/2}=   a Z_c Z_d \sqrt{\dfrac{\mathcal{A}}{Q}} + b \sqrt{ \mathcal{A} (A_c^{1/3}+A_d^{1/3}) Z_c Z_d } +c \dfrac{l(l+1)}{ \sqrt{ \mathcal{A} (A_c^{1/3}+A_d^{1/3}) Z_c Z_d }} + d 
\end{equation}

\end{widetext}%

where the constant coefficients $b$ and $d$ are different since now they also include additional contributions due to the formation probability. This is the most general formulation of Geiger-Nuttall law called the universal decay law of which the simple Geiger-Nuttall law [Eq. \eqref{GN}] and its generalizations (e.g., UF and Viola-Seaborg law) are special cases. It was claimed by their authors to be valid for all nuclei and decay modes \cite{UDL1, UDL2, GNreview, GNreview2}. The validity of this law or the simple or any generalized Geiger-Nuttall laws hinges on the validity of the assumption that $\gamma_l(x)$ can be linearized. It must be emphasized that this assumption is not unique to the R-matrix theory and is required in any derivation of Geiger-Nuttall law (simple or generalized) since the term $\exp[- 2\gamma_l(x)]$ proportional to the Coulomb penetration is present in all models (e.g., see Refs. \cite{UF} and \cite{surfacewell}). In all derivations of Geiger-Nuttall laws in the literature, the linearity assumption is justified based on the claim that $x$ should always be small, i.e., $x_0=0$ and consequently we have $a_0=\pi/2$ [Eq. \eqref{a}] and $b_0=-2$ [Eq. \eqref{b}]. This was first claimed by Gamow \cite{gamowx} and all subsequent authors echoed the same claim in their derivations \cite{xapp0, xapp1, xapp2, xapp3, UF, UDL1, UDL2, radius, brown}. In particular, the authors of the UDL argued that $x$ should become progressively smaller for heavier cluster-daughter systems (i.e., large $Z_c Z_d$) since the denominator increases [see Eq. \eqref{x}]. Consequently, the linearity assumption and the UDL should become more exact \cite{UDL1, UDL2}. However, the numerator of $x$ [Eq. \eqref{x}] has $R$ and $Q$ which also increase with heavier cluster-daughter systems and therefore, it would be premature to conclude that $x$ becomes smaller. In the next section, we systematically study the variation of $x$ across nuclei and decay modes and its implications for the validity of the three classes of the Geiger-Nuttall laws.

\section{results and discussion} \label{results}

\subsection{Variation of $x$ in the heavy and superheavy region} \label{xvar}

 \begin{figure*}[htp!]
 \centering

  \includegraphics[width= \textwidth,keepaspectratio]{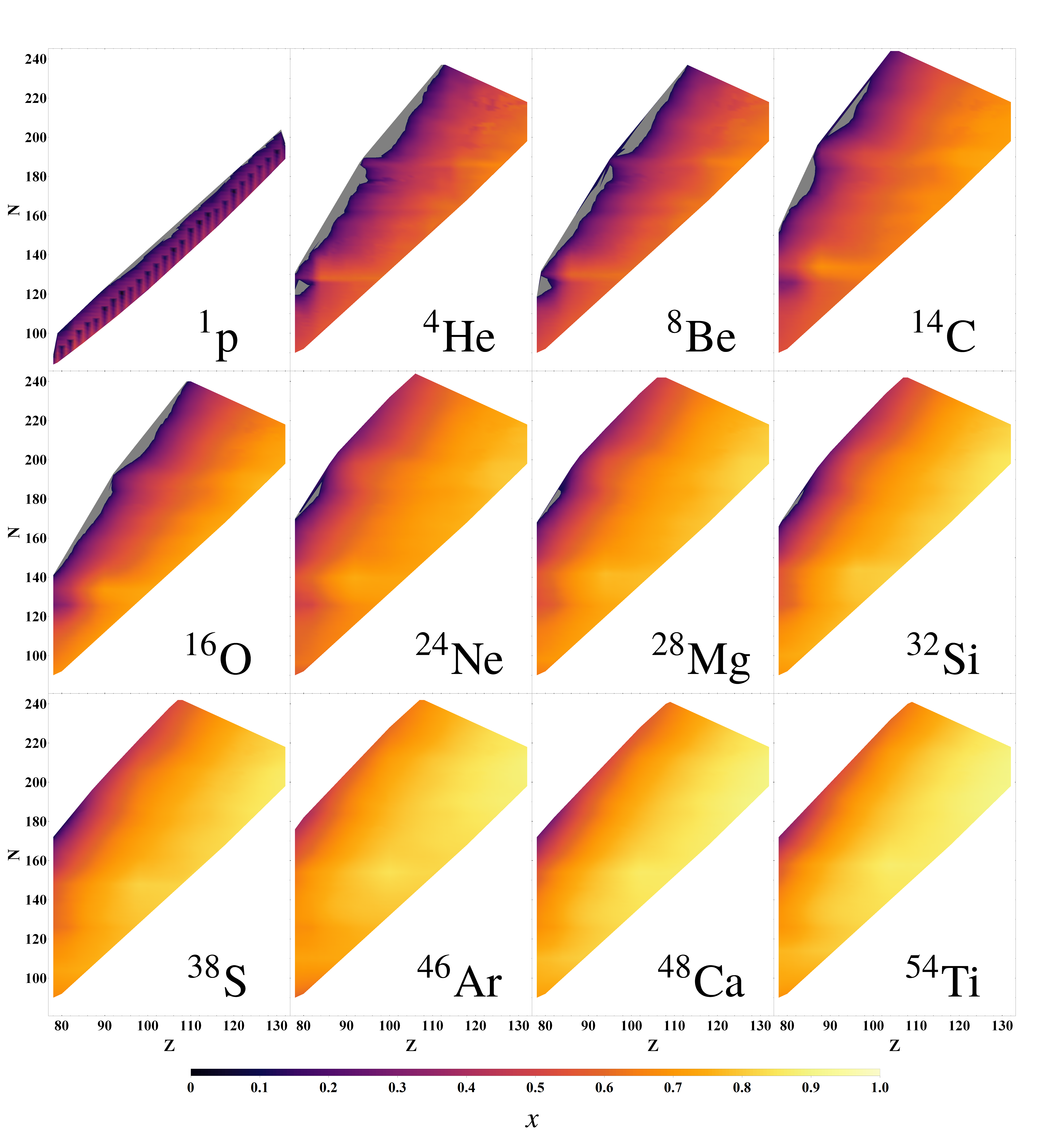}%

      \caption{ Density plot of $x$ vs parent proton and neutron numbers $Z$ and $N$ for clusters from $\isotope[1]{p}$ up to $\isotope[54]{Ti}$.}
        \label{ZNx1}

\end{figure*}





To be able to reach a general and rigorous conclusion about the linearity assumption for all nuclei and decay modes, we systematically investigated 5130 nuclei in the region $162 \le A \le 350 $ and $78 \le Z \le 132$. For each cluster-daughter system considered, we calculated its $x$ using Eq. \eqref{x}. We chose 12 cluster decay modes ranging from $\isotope[1]{p}$ up to $\isotope[54]{Ti}$. $Q$ in $x$  [Eq. \eqref{x}] is given by 

\begin{equation} \label{Q}
Q= M_p -(M_c +M_d)
\end{equation}

where $M_p, M_c$, and $M_d$ are the excess masses of the parent, cluster, and daughter nuclei, respectively. Experimental mass was used and when not available, we used the theoretical values computed by the WS4 model with the radial basis function (RBF) correction or WS$4^{\text{RBF}}$  \cite{WS4,WS4Table}. This model was chosen since it is the most accurate mass model in reproducing experimental masses and decay energies for heavy and superheavy nuclei with root-mean-square (rms) deviation of 170 KeV from experimental mass excess \cite{WS4, TQstudy, Mcompare}. Moreover, systematic studies and recent new mass measurements (e.g., $\isotope[249,250,252]{Md},\isotope[253,254]{Lr}, \isotope[257,258]{Db},\isotope[261,262]{Bh}$, and $\isotope[266]{Mt}$) show that it has the best predictive power compared to other models \cite{Mcompare, WSpower, newM}.

The result of our calculations is shown in Fig. \ref{ZNx1} as a density plot of $x$ vs parent proton and neutron numbers $Z$ (horizontal axis) and $N$ (vertical axis).  Each subfigure shows the variation of $x$ for a given decay mode across the nuclear chart. Darker colors denote smaller $x$ and vice versa (see the bottom of Fig. \ref{ZNx1} for the color scheme) while the gray color denotes regions in which $x$ is not defined because $Q$ is negative (i.e., nuclei stable against decay). Regions with negative proton and neutron separation energies were excluded in the density plots of $\isotope[4]{He}$ to $\isotope[54]{Ti}$ since they are unbound nuclei. Contrary to the expectation of previous researchers that $x$ is always close to zero, $x$ varies over the whole range of its possible values between 0 and 1 with a minimum of 0.004 and a maximum of 0.904. The average value of $x$ progressively increases with larger decay modes from 0.26 for proton decay, 0.43 in $\alpha$ decay up to 0.78 for $\isotope[54]{Ti}$ decay. The standard deviation from the average is about 0.11 in a given decay mode. This increase in the average $x$ is owing to the average $Q$ progressively increasing from 1.6 MeV for proton decay, 6.4 MeV for $\alpha$ decay up to 130 MeV for $\isotope[54]{Ti}$ decay. Thus in large cluster-daughter systems, although the denominator of $x$ [Eq. \eqref{x}] (i.e., $Z_c Z_d$) increases, the numerator (i.e., $QR$) increases at a faster rate such that $x$ increases overall. In a given decay mode, the figure shows that $x$ is the largest in proton-rich nuclei with large $Z$ and small $N$ and smallest in neutron-rich nuclei with a small $Z$ and large $N$. For a fixed $Z$, $x$ progressively decreases as $N$ increases (i.e., move vertically upwards in the subfigure). For a fixed $N$ on the other hand, $x$ progressively increases as $Z$ increases (i.e., move horizontally to the right in the subfigure). These patterns which hold consistently in any given decay mode are due to $Q$ increasing for proton-rich nuclei while decreasing for neutron-rich ones.

\subsection{Implications for the universal Geiger-Nuttall law} \label{UGN}

 We can sum up all the above points in the following three observations:

\begin{enumerate} 

\item $x$ varies over its whole range between $0$ and $1$ when all nuclei and decay modes are considered.
\item There is a very large difference in the $x$ values between the lighter and heavier decay modes where $x$ is the largest in the latter.
\item In a given decay mode, there is a less large but still significant difference in the $x$ values between the neutron-rich and proton-rich nuclei where $x$ is the largest in the latter. 
\end{enumerate}

These three observations (especially the first) lead us to the first major conclusion in the paper: a universal Geiger-Nuttall law encompassing all nuclei and decay modes with a fixed set of coefficients is not possible. This is the case since the linearity assumption which requires that $x$ varies within an interval is manifestly false when $x$ varies over the whole range. To gain a deeper understanding of the validity of the simple, generalized, and universal Geiger-Nuttall laws in relation to the linearity assumption, define $\sigma(x)$ which is proportional to $\log_{10}T_{1/2}$ [Eq. \eqref{logT}] as

\begin{equation} \label{DelAlphaExp}
\sigma(x)=\dfrac{\gamma_0(x)}{\eta}=  \arccos(x) -x \sqrt{1-x^2}
\end{equation}

\begin{figure}
 \centering

  \includegraphics[width= 0.5\textwidth,keepaspectratio]{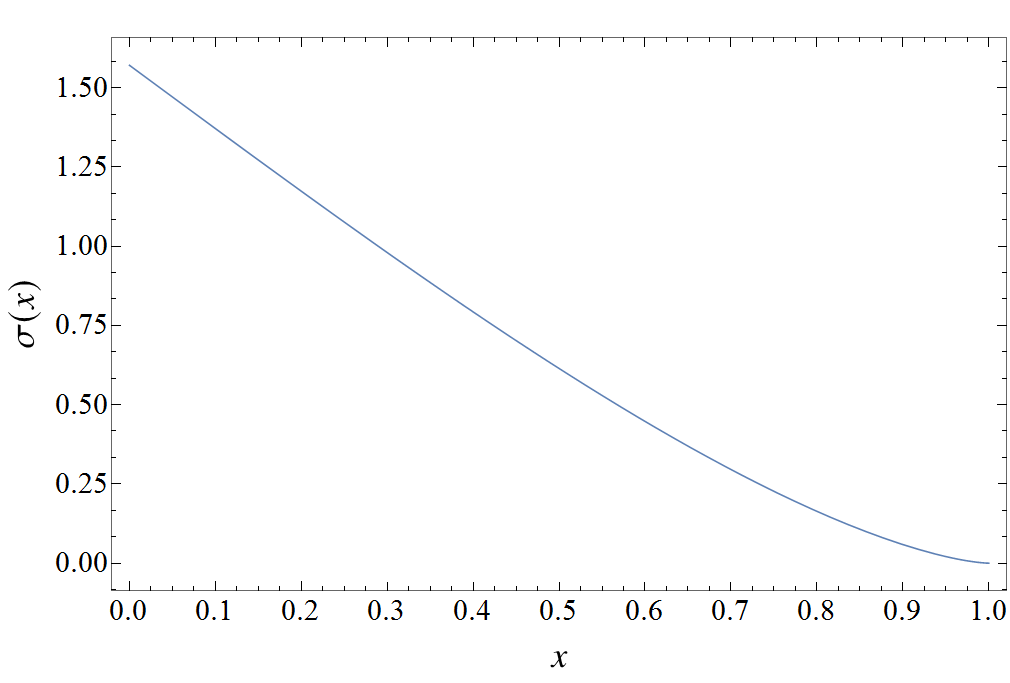}%

      \caption{ $\sigma(x)$ [Eq. \eqref{sigma}] vs $x$.}
        \label{sigma}

\end{figure}

\begin{figure}
 \centering

  \includegraphics[width= 0.5\textwidth,keepaspectratio]{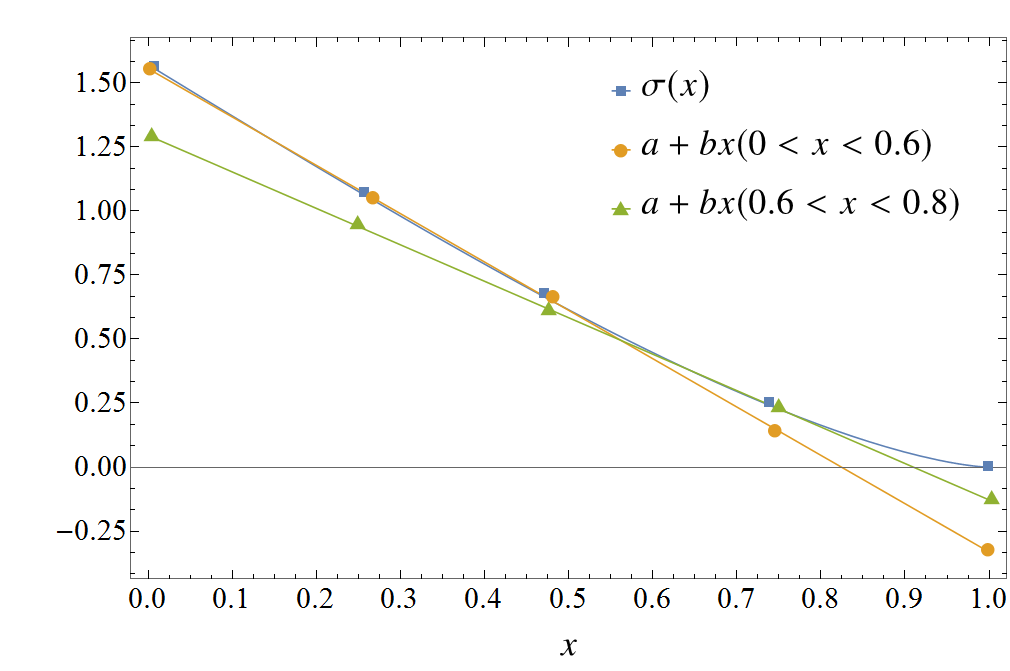}%

      \caption{Two best linear fits in $0<x<0.6$ (orange circle) and $0.6<x<0.8$ (green triangle) for $\sigma(x)$ (blue square). }
        \label{UGNfig}

\end{figure}

The simple Geiger-Nuttall law and its generalizations do not hold unless $x$ varies in an interval where $\sigma(x)$ can be approximated as a linear function of $x$. $\sigma(x)$ is plotted in Fig. \ref{sigma} where it is approximately linear over more than half its range and starts to become noticeably non-linear from about $x \ge 0.8$. In Fig. \ref{UGNfig}, we show two best linear fits $a+bx$ for $\sigma(x)$ (blue square) where one is fitted for the interval $0<x<0.6$ (orange circle) and the other for $0.6<x<0.8$ (green triangle). Evidently, a universal Geiger-Nuttall law is not possible since there is no single straight line that can approximate $\sigma(x)$ over the whole interval. There is also a fundamental trade-off where accurately approximating $x$ over an interval comes at the expense of large errors for regions outside the fit, e.g., both linear fits in Fig. \ref{UGNfig} strongly deviate from $\sigma(x)$ outside their interval of the fitting. This implies that an accurate description of light decay modes and neutron-rich nuclei (i.e., small $x$) comes at the expense of large errors in heavy clusters and proton-rich nuclei (i.e., large $x$) and vice versa. As will be discussed shortly, for $x\ge 0.8$ in particular, $\sigma(x)$ starts to become non-linear such that no reasonable linear approximation can accurately capture its behavior in this interval. Next, we show how the variation of $x$ explains the success of the simple Geiger-Nuttall law in $\alpha$ decay.  

\subsection{The validity of the simple and generalized Geiger-Nuttall laws in light cluster decay} \label{GNvalid}

In the case of $\alpha$ decay with $78 \le Z \le 110$, all the $x$ values of the 520 experimentally observed nuclei lie in $0 < x < 0.6$ (with most points lying around $x=0.5$) where $\sigma(x)$ can be reasonably represented as a straight line. This explains the reason for the past success of the simple Geiger-Nuttall law in $\alpha$ decay. Moreover, the reason for the success of generalizations like the Viola-Seaborg law in unifying different isotopic chains is that $\sigma(x)$ can be described with a single straight line $a+b x$ for $0 < x < 0.6$. The error $\Delta$ in $\log_{10}T_{1/2}$ [Eq. \eqref{logT}] from the linearization $a+bx$ is given by

\begin{equation} \label{Del}
\Delta=\dfrac{2 \eta}{\ln 10}[\sigma(x) -(a+bx)]
\end{equation} 

 \begin{figure}
 \centering

  \includegraphics[width= 0.5\textwidth,keepaspectratio]{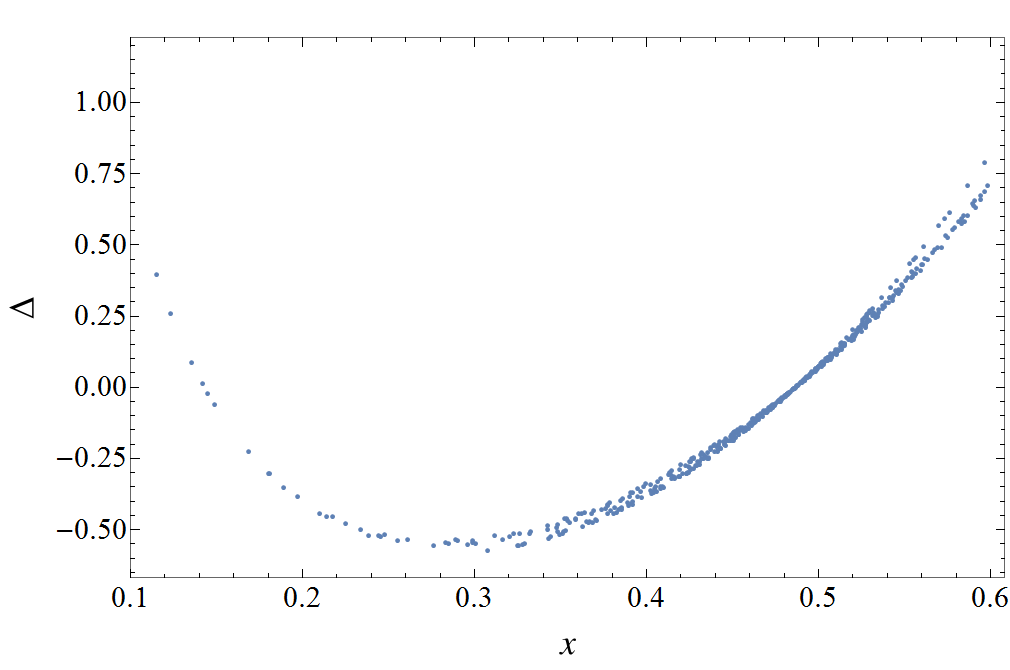}%

      \caption{ $\Delta$ [Eq. \eqref{Del}] vs $x$ for the best linear fit of $\sigma(x)$ in $0<x<0.6$.}
        \label{DelAlphaExp}

\end{figure} 

In Fig. \ref{DelAlphaExp}, $\Delta$ resulting from using a single best-fitted straight line $a+bx$ in $0 < x < 0.6$ for 520 nuclei decaying by $\alpha$ emission is shown. $\Delta$ is less than one order of magnitude for all nuclei (except for four nuclei with $x<0.1$ not shown in this figure) with the vast majority of nuclei restricted to $0.4<x<0.5$ and $|\Delta|<0.5$. The resulting rms deviation in $\log_{10}T_{1/2}$ is 0.33 which is within the typical values for the simple and generalized Geiger-Nuttall laws in $\alpha$ decay. Looking at the extrapolations of $x$ across the nuclear chart in the density plot of $\isotope[4]{He}$ in Fig. \ref{ZNx1}, we see that the vast majority of nuclei have $x<0.6$ and thus the linearity assumption is true. Consequently, we expect the simple Geiger-Nuttall law and its generalizations to provide an accurate description of $\alpha$ decay of all isotopic chains that are to be observed in future experiments. Similar conclusions hold for the lighter cluster decay mode of proton.

\begin{figure}
 \centering

  \includegraphics[width= 0.5\textwidth,keepaspectratio]{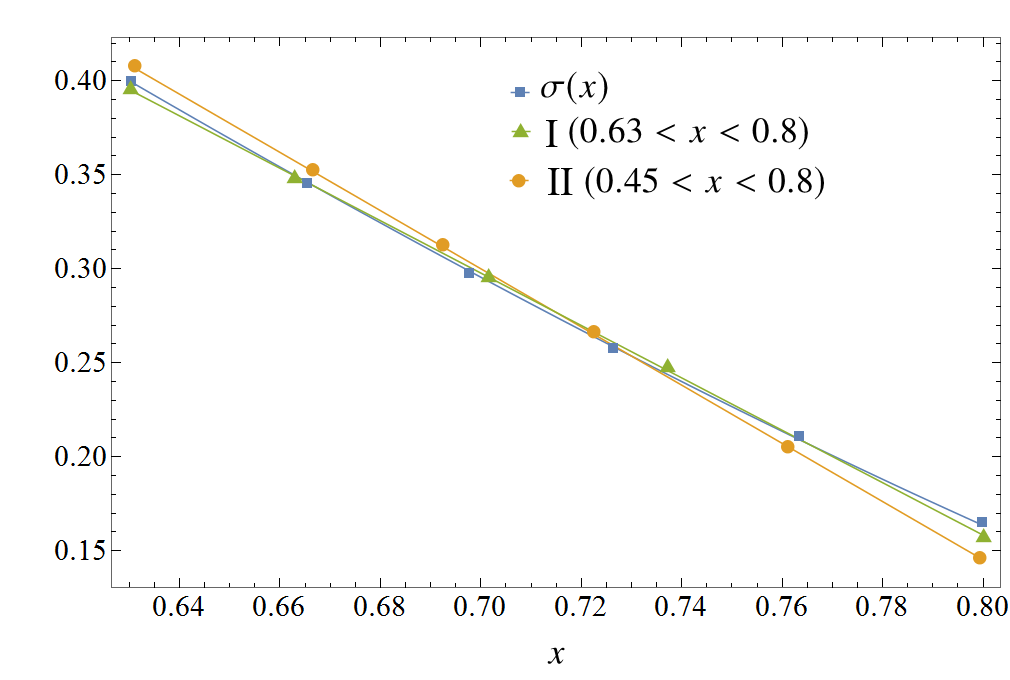}%

      \caption{best linear fit of $\sigma(x)$ (blue square) for $\alpha$ and cluster decay with $0.45<x<0.8$ (orange circle) and best fit for cluster decay with $0.63<x<0.8$ (green triangle).}
        \label{GGN}

\end{figure}

Similarly, the insights about the linearity assumption can shed light on the limitations of generalized Geiger-Nuttall laws encompassing different nuclei and more than a decay mode. In Table \ref{xtable}, we show the $x$ values of 11 trans-lead nuclei decaying by cluster emission in increasing order of $x$. $\log_{10} T^{\text{exp}}_{1/2}$ denotes the experimental half-life. Their $x$ values are between 0.63 and 0.8, larger than the maximum $x$ in $\alpha$ decay of the 520 nuclei just investigated. There is a direct correlation between the cluster-daughter size and $Q$ on the one hand and $x$ on the other where they increase with each other. In the table, we consider UDL as an example of a generalized Geiger-Nuttall law---being the most general. $\log_{10} T^{\text{I}}_{1/2}$ denotes the half-life calculated by the UDL fitted for cluster decay data alone. $\log_{10} T^{\text{II}}_{1/2}$ denotes the half-life calculated by the UDL formula fitted for $\alpha$ and cluster decay experimental data. The fitted coefficients $a, b$ and $d$ in Eq. \eqref{UGN2} ($c=0$ for even-even nuclei with $l=0$) are taken from Refs. \cite{UDL1, UDL2}. While $\log_{10} T^{\text{I}}_{1/2}$ reproduces $\log_{10} T^{\text{exp}}_{1/2}$ well, there are two interesting anomalies related to $\log_{10} T^{\text{II}}_{1/2}$ in need of explanation. First, $\log_{10} T^{\text{II}}_{1/2}$ is very poor in reproducing experimental half-life. The average error is one order of magnitude and can be as large as 2.2. Second, $\log_{10} T^{\text{II}}_{1/2}$ is significantly and systematically larger than $\log_{10} T^{\text{I}}_{1/2}$ (and $\log_{10} T^{\text{exp}}_{1/2}$) for the small clusters while it is significantly smaller than $\log_{10} T^{\text{I}}_{1/2}$ for the large clusters. Both of these facts can be explained through our ongoing discussion about the variation of $x$ with the decay modes and the linearization of $\sigma(x)$. In Fig. \ref{GGN}, $\sigma(x)$ (blue square) is shown alongside with its best linear fit in the interval $0.45<x<0.8$ (orange circle, line II) corresponding to $\alpha$ and cluster decay and the best linear fit in the interval $0.63<x<0.8$ (green triangle, line I) corresponding to cluster decay alone. The choice of the particular interval $0.45<x<0.8$ here is only for the illustrative purpose of showing the effect of describing $\alpha$ decay together with heavier clusters. From the figure, we see that line II (proportional to $\log_{10} T^{\text{II}}_{1/2}$) is worse than line I (proportional to $\log_{10} T^{\text{I}}_{1/2}$) in approximating $\sigma(x)$ (proportional to $\log_{10} T^{\text{exp}}_{1/2}$) in $0.63<x<0.8$. This explains the first fact related to the large deviation of  $\log_{10} T^{\text{II}}_{1/2}$ from experimental half-life. Second, line II is larger than line I [and $\sigma(x)$] for smaller $x$ with the largest gap for the smallest $x=0.63$ (i.e., smallest cluster size $\isotope[14]{C}$). The two lines begin to converge as $x$ increases until they intersect at $x \approx 0.72$. Then they diverge again where line I becomes progressively larger than line II for all larger $x$. The gap between them becomes the largest at the largest $x=0.8$ (i.e., largest cluster size $\isotope[34]{Si}$). We see this exact pattern mirrored in the table as we move down the rows corresponding to larger $x$ and cluster size. We see $\log_{10} T^{\text{II}}_{1/2}-\log_{10} T^{\text{I}}_{1/2}=2.61$ is the largest for the smallest $x=0.63$ (i.e., first row $\isotope[226]{Ra} \to \isotope[14]{C}$) and the difference approaches zero for larger $x$ until $\log_{10} T^{\text{II}}_{1/2}-\log_{10} T^{\text{I}}_{1/2}=0.25$ at $x=0.75$ (i.e., $\isotope[232]{U}\to \isotope[24]{Ne}$). Afterwards, the difference becomes negative reaching a minimum at $x=0.8$ of $\log_{10} T^{\text{II}}_{1/2}-\log_{10} T^{\text{I}}_{1/2}=-1.95$ (i.e., last row $\isotope[242]{Cm} \to \isotope[34]{Si}$). 

Therefore, we learn from the analysis in this subsection that while it is possible to describe $\alpha$ ($0<x<0.6$) and heavier cluster decays ($x>0.6$) separately using two distinct straight lines with reasonable accuracy (see Figs. \ref{UGNfig} and \ref{GGN}), attempting to unify them with a single straight line (e.g., $\log_{10} T^{\text{II}}_{1/2}$) results in unacceptable errors since $x$ varies over a wide range from 0 to 0.8 such that no single straight line can accurately describe $\sigma(x)$ (see Fig. \ref{UGNfig}). This establishes fundamental theoretical limitations for unifying $\alpha$ with heavier cluster decay modes for all generalized Geiger-Nuttall laws. We conclude this subsection by stressing that the linearity assumption is a necessary but not a sufficient condition to ensure the validity of Geiger-Nuttall law. The formation probability $\log_{10} |R g_l(R)|^2$ also has to vary smoothly in Eq. \eqref{logT}. While this is generally the case in the experimental region investigated \cite{GNreview,GNreview2, UDL1, UDL2}, there are exceptions (e.g., abrupt decline at shell closures like $N=126$) which result in $\log_{10}T_{1/2}$ deviating from the predictions of the generalized Geiger-Nuttall laws by more than one order of magnitude \cite{UNIVUDL} (for discussion about the validity of the simple Geiger-Nuttall law in connection to the formation probability $\log_{10} |R g_l(R)|^2$, see Ref. \cite{GNvalid}). 

\begin{table}[htp!]
\caption{The half-life calculated by the UDL fitted for $\alpha$ and cluster decay data  $\log_{10} T^{\text{II}}_{1/2}$  vs the one fitted for cluster decay alone $\log_{10} T^{\text{I}}_{1/2}$. Experimental half-life $\log_{10} T^{\text{exp}}_{1/2}$ is taken from Ref. \cite{Texp}. Corresponding $x$ of every decay is also shown. } 
\label{xtable}
\begin{ruledtabular}
\begin{tabular}{c c c c c c c}
Parent   &  Cluster   & $Q$ (MeV)     &  $\log_{10} T^{\text{exp}}_{1/2}$ & $\log_{10} T^{\text{II}}_{1/2}$ & $\log_{10} T^{\text{I}}_{1/2}$ & $x$ \\ \hline
$\isotope[226]{Ra}$ & $\isotope[14]{C}$ & 28.20 & 21.29 & 22.93 & 20.32   & 0.63 \\
$\isotope[224]{Ra}$ & $\isotope[14]{C}$ & 30.54 & 15.90 & 17.83 & 15.58  & 0.66 \\
$\isotope[222]{Ra}$ & $\isotope[14]{C}$ & 33.05 & 11.05 & 12.99  & 11.07 & 0.68   \\
$\isotope[228]{Th}$ & $\isotope[20]{O}$ & 44.72 & 20.73 & 22.95 & 21.56   & 0.70 \\
$\isotope[230]{Th}$ & $\isotope[24]{Ne}$ & 57.76 & 24.61 & 25.38 & 24.74   & 0.73 \\
$\isotope[230]{U}$ & $\isotope[22]{Ne}$ & 61.39 & 19.56 & 21.23 & 20.73   & 0.74 \\
$\isotope[232]{U}$ & $\isotope[24]{Ne}$ & 62.31 & 20.39 & 20.93 & 20.68   & 0.75 \\
$\isotope[234]{U}$ & $\isotope[28]{Mg}$ & 74.11 & 25.74 & 24.99 & 25.36   & 0.76 \\
$\isotope[236]{Pu}$ & $\isotope[28]{Mg}$ & 79.67 & 21.65 & 20.23 & 21.02   & 0.78 \\
$\isotope[238]{Pu}$ & $\isotope[32]{Si}$ & 91.19 & 25.30 & 23.99 & 25.39   & 0.78 \\
$\isotope[242]{Cm}$ & $\isotope[34]{Si}$ & 96.51 & 23.11 & 20.92 & 22.87   & 0.8 \\

\end{tabular}
\end{ruledtabular}
\end{table}

\subsection{Discrepancies and extrapolative limitations of generalized Geiger-Nuttall laws in heavy cluster decay} \label{GNpredict}

It is instructive to explain the discrepancies in half-life predictions of heavy cluster decay arising from using different generalized Geiger-Nuttall laws, namely, the UDL, UF, and Horoi formula. For even-even nuclei (i.e., $l=0$), they are given by \cite{UDL1,UDL2, UF, horoi}  

\begin{equation} \label{UDL}
\log_{10}T^{\text{UDL}}_{1/2} =   a_1 Z_c Z_d \sqrt{\dfrac{\mathcal{A}}{Q}} + a_2 \sqrt{ \mathcal{A} (A_c^{1/3}+A_d^{1/3}) Z_c Z_d } + a_3 
\end{equation}

\begin{equation} \label{UF}
\log_{10}T^{\text{UF}}_{1/2} =   a_1 Z_c Z_d \sqrt{\dfrac{\mathcal{A}}{Q}} + a_2 \sqrt{ \mathcal{A}  Z_c Z_d } +a_3
\end{equation}

\begin{equation} \label{Horoi}
\log_{10}T^{\text{Horoi}}_{1/2}=   (a_1 \sqrt{\mathcal{A}}+a_2) [ \dfrac{(Z_c Z_d)^{0.6}}{\sqrt{Q}}-7]
  +  (a_3 \sqrt{\mathcal{A}}+a_4)
\end{equation}

\begin{figure}
 \centering

  \includegraphics[width= 0.5\textwidth,keepaspectratio]{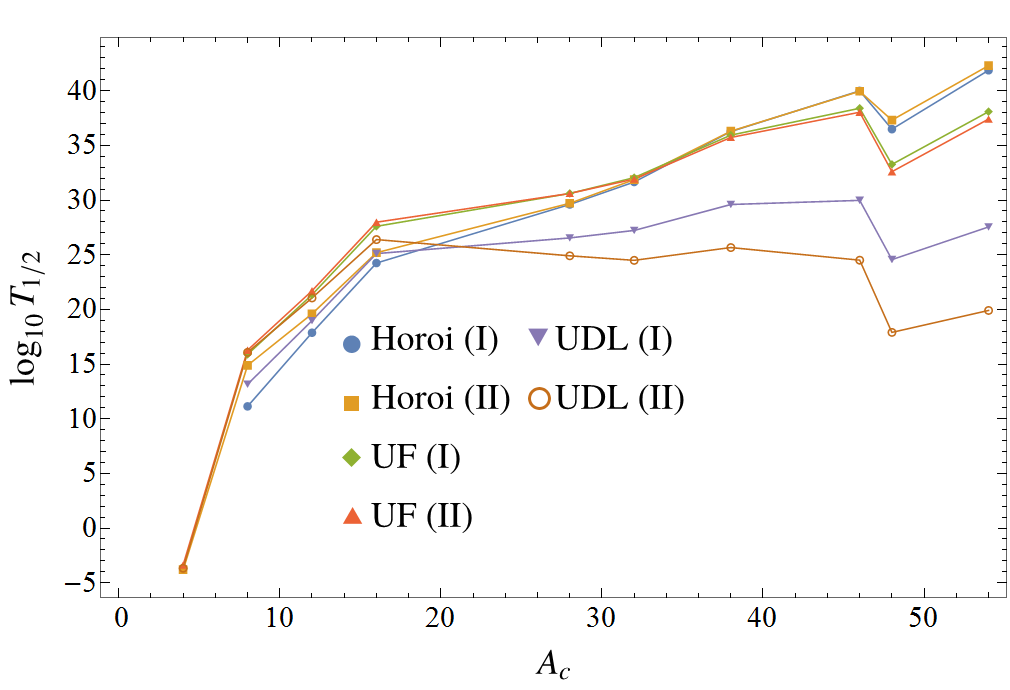}%

      \caption{Half-life predictions $\log_{10}T_{1/2}$ by three generalized Geiger-Nuttall laws of various cluster decays vs $A_c$ for the parent nuclei $^{294}{118}$.}
        \label{UFandHoroiandUDL}

\end{figure}

In the UDL, the first two terms are proportional to $\eta$ and $\eta x$, respectively, as have been shown in Sec. \ref{model}. UF is identical to UDL with the only difference that $R=r_0(A_c^{1/3}+A_d^{1/3})$ in the second term is approximated as a constant for all decay modes and nuclei, i.e., $x$ is approximated as $\sqrt{R_0 Q/(e^2 Z_c Z_d)}$ where $R_0$ is a constant implicitly determined by fitting $a_2$ in Eq. \eqref{UF}. The Horoi formula differs in its functional form from the other two in that it exhibits a $(Z_c Z_d)^{0.6}Q^{-1/2}$ dependence instead of $Z_c Z_d Q^{-1/2}$ and lacks the $\eta x$ term (second term in the UDL and UF). While the UDL and UF can be derived from the R-matrix theory, the Horoi formula lacks a rigorous theoretical basis and is justified by fitting to the experimental data [also the $(Z_c Z_d)^{0.6}$ dependence is not universal for all decay modes and becomes $(Z_c Z_d)^{0.8}$ for proton emission] \cite{brown, horoi,brownp}. Systematic studies of heavy cluster radioactivity show discrepancies in half-life predictions between these three formulas larger than 60 orders of magnitude \cite{GNpredict1, GNpredict2, GNpredict3}. Moreover, there is a consistent pattern in which the Horoi formula systematically predicts the largest half-life for a given decay followed by the UF then UDL \cite{GNpredict1, GNpredict2, GNpredict3}. The present findings of variation of $x$ with decay modes coupled with how $x$ is incorporated differently in the above equations can provide a plausible explanation of the discrepancies just mentioned. Consider as an illustrative example, ten clusters from $\isotope[4]{He}$ up to $\isotope[54]{Ti}$ emitted from the parent nuclei $^{294}{118}$. $Q$ and the corresponding $x$ value of each decay are shown in Table \ref{xtable2}. The corresponding half-life predictions of the three formulas vs cluster mass number $A_c$ are shown in Fig. \ref{UFandHoroiandUDL}. For each of the three formulas, we predicted half-life using the coefficients fitted for experimental $\alpha$ and cluster decay data (dubbed II in Fig. \ref{UFandHoroiandUDL}) and the coefficients fitted for experimental cluster decay data alone (dubbed I) which are taken from Refs. \cite{UDL1,UDL2, UF, horoi}. For a given parent nuclei, $\log_{10} T_{1/2}$ depends on the product of two opposing influences $\eta \sigma(x)$ where on average $\eta$ increases with cluster size while $\sigma(x)$ decreases (since $x$ increases). While UDL and UF incorporate such dependence approximately as $\eta(a+bx)$ [first two terms in Eqs. \eqref{UDL} and \eqref{UF}], it is lacking in the Horoi formula which only incorporates the $\eta$ dependence [more precisely, $(Z_c Z_d)^{-0.4} \eta$]. This explains why $\log_{10}T^{\text{Horoi}}_{1/2}$ is systematically larger than the predictions of the other two formulas in heavy clusters since it does not incorporate $\sigma(x)$ which decreases with cluster size. Indeed, Fig. \ref{UFandHoroiandUDL} shows that $\log_{10}T^{\text{Horoi}}_{1/2}$ is significantly less sensitive to the change in $x$ between $A_c=46$ and $54$ (see Table \ref{xtable2}) compared to the other two formulas.

That $\log_{10}T^{\text{UF}}_{1/2}$ is systematically greater than $\log_{10}T^{\text{UDL}}_{1/2}$ is a consequence of UF approximating $R=r_0(A_c^{1/3}+A_d^{1/3})$ (which increases with cluster mass) as constant for all decay modes and thus systematically underestimating $x$ [i.e., overestimating $\sigma(x)$] compared to UDL. Another fact concerning UDL and UF is that predictions of the formulas fitted for $\alpha$ and cluster data are systematically smaller than those fitted for cluster decay data alone for large $A_c$. This is the case because both $a$ and $|b|$ ($b$ is always negative) in the linearization $a+bx$ decrease when the average $x$ involved in the fitting increases (i.e., heavier decay modes) since $a_0$ [Eq. \eqref{a}] and $b_0$ [Eq. \eqref{b}] parametrizing the tangent line to $\sigma(x)$ at $x=x_0$ monotonically decrease with $x_0$.  Indeed $a_1$ decreases with heavier cluster decay modes (i.e., larger average $x$) in UDL, where $a_1$ in Eq. \eqref{UDL} is 0.4065 for data fitted for $\alpha$ decay data alone, 0.3949 for $\alpha$ and cluster decay data, and 0.3671 for cluster decay data alone \cite{UDL1,UDL2}. Similarly, $a_2$ is -0.4311, -0.3693, and -0.3296, respectively \cite{UDL1, UDL2}. The same pattern holds true for the coefficients of UF \cite{UF}. Now a linearization of $\sigma(x)$ with larger intercept $a$ and slope $b$ would be systematically smaller than a linearization with smaller $a$ and $b$ for large $x$ (e.g., see Figs. \ref{UGNfig} and \ref{GGN}). Consequently, UDL (and UF) fitted for $\alpha$ and cluster decay data (i.e., larger $a$ and $b$) is systematically smaller than the one fitted for cluster data alone for large $A_c$ or $x$. This is especially pronounced in UDL and makes it numerically unstable and unreliable where a difference of $10^{-2}$ between $a_1$ fitted for $\alpha$ and cluster data vs cluster data alone (same for $a_2$) translates into 13 orders of magnitude of discrepancy in their half-life predictions for heavy clusters, i.e., $\log_{10}T^{\text{UDL}}_{1/2}$ is very sensitive to the fitting parameters $a_1$ and $a_2$ (see Fig. \ref{UFandHoroiandUDL}). Interestingly, the Horoi formula fitted for cluster decay data alone is systematically smaller than the one fitted for $\alpha$ and cluster decay data unlike UF and UDL. However, the same line of reasoning cannot be applied on the Horoi formula to explain this systematic deviation since it does not exhibit the $a+bx$ dependence we are discussing.

The important conclusion is that all of these generalized Geiger-Nuttall laws lack any extrapolative power for heavy cluster decay predictions (i.e., $x>0.8$ or clusters heavier than $\isotope[34]{Si}$) since they were originally fitted for $x<0.8$ and such linearizations strongly deviate from $\sigma(x)$ beyond $x=0.8$ (see Fig. \ref{UGNfig}). Next, we will show that there is an even more fundamental problem, namely, that a Geiger-Nuttall description of clusters that includes heavy ones (i.e., $x>0.8$) must breakdown due to non-linearity.

\begin{table}[htp!]
\caption{ $Q$ (computed by WS$4^{\text{RBF}}$) and $x$ of various cluster decays from the parent nuclei $^{294}{118}$. } 
\label{xtable2}
\begin{ruledtabular}
\begin{tabular}{c c c c c c c}
 Cluster   & $Q$ (MeV)     & $x$ & Cluster & $Q$ (MeV) & $x$ \\ \hline
$\isotope[4]{He}$ & 12.198 & 0.600 &  $\isotope[32]{Si}$ & 119.316 & 0.809 \\
$\isotope[8]{Be}$ & 23.191 & 0.603 &   $\isotope[38]{S}$ &  136.880 &  0.824   \\
$\isotope[12]{C}$ & 40.527 &  0.667 &  $\isotope[46]{Ar}$ &  155.717 & 0.843 \\
$\isotope[16]{O}$ & 57.459 & 0.702 &   $\isotope[48]{Ca}$  & 177.976 &   0.866 \\
$\isotope[28]{Mg}$ & 100.134 & 0.788 & $\isotope[54]{Ti}$  &  191.737 & 0.870  \\

\end{tabular}
\end{ruledtabular}
\end{table}

\subsection{Nonlinearity and breakdown of Geiger-Nuttall description in heavy cluster decay} \label{Nlaw}

\begin{figure}
 \centering

  \includegraphics[width= 0.5\textwidth,keepaspectratio]{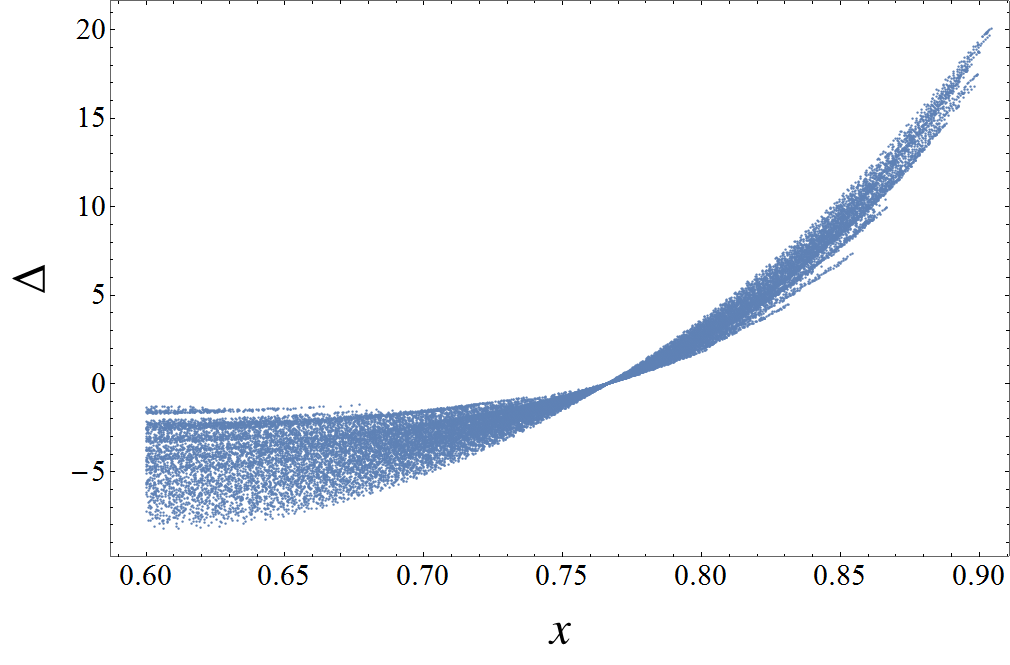}%

      \caption{ $\Delta$ [Eq. \eqref{Del}] from the best linear fit in $0.6<x<1$ for all nuclei and decay modes.}
        \label{GGNError}

\end{figure}

We have already seen how $\sigma(x)$ is non-linear in the sense that its slope of the tangent continuously changes with $x$ thus preventing an accurate linear approximation of $\sigma(x)$ over big intervals (see Figs. \ref{sigma} and \ref{GGN}). There are two other facts which exacerbate the problem even further particularly for heavier cluster decay modes. First, a linear approximation of $\sigma(x)$ for an interval of a fixed length becomes progressively worse for larger $x$, e.g., the rms error of a linear fit of $\sigma(x)$ in the interval $0<x<0.2$ is 0.0003, 0.003 in $0.6<x<0.8$, and 0.0068 in $0.8<x<1$. Second, the error in half-life from linearization $\Delta$ [Eq. \eqref{Del}] depends on the product of two factors, namely, 1) the error in approximating $\sigma(x)$ by $a+bx$ (i.e., their difference) and 2) $2 \eta / \ln 10=0.868 \eta$. The Coulomb parameter $\eta$ is in the same order of magnitude for a given decay mode and becomes progressively larger with larger clusters, e.g., the average $\eta$ is 67.6 in $\alpha$ decay, 160.4 in $\isotope[14]{C}$ decay, and 260 in $\isotope[32]{Si}$ decay [calculated using Eq. \eqref{eta2} and the experimental $Q$ or WS$4^{\text{RBF}}$]. Thus even small errors resulting from the linearization of $\sigma(x)$ get multiplied by the large factor $0.868 \eta$ translating into significant deviations from $\log_{10} T_{1/2}$. For instance, for the decay $\isotope[226]{Ra} \to \isotope[14]{C}$ with $\eta=105.75$ (see Table \ref{xtable}), an error of 0.01 due to linearization translates to about one order of magnitude of error in $\log_{10} T_{1/2}$ ($0.868 \eta \times 0.01=0.92$). The same error of 0.01 due to linearization in the decay $\isotope[242]{Cm} \to \isotope[34]{Si}$ with $\eta=198.97$ translates to about two orders of magnitude of error in $\log_{10} T_{1/2}$ ($0.868 \eta \times 0.01=1.72$). All of these effects we just described are especially pronounced in heavier cluster decay modes (i.e., large $x$) where $\eta$ is even larger and $\sigma(x)$ is poorly approximated as a linear function. In Fig. \ref{GGNError}, we show the error $\Delta$ resulting from the best linear fit of $\sigma(x)$ for all nuclei across the nuclear chart and all decay modes with $0.6<x<1$. The figure marks the breakdown of Geiger-Nuttall law description of heavy cluster decay modes. It shows that $\sigma(x)$ cannot be reasonably approximated linearly in the interval $0.6<x<1$. It also establishes the effect of the multiplicative factor $0.868\eta$ in translating the errors (due to linearization) into errors in $\log_{10}T_{1/2}$ larger than 15 orders of magnitude. We conclude that a generalized Geiger-Nuttall law encompassing all cluster decay modes is not possible contrary to the claims of previous authors (e.g., authors of the UDL and UF) \cite{UDL1, UDL2, UF}. 





Moreover, as discussed before, a direct implication of the fact that the ratio $H_{l}^{+}(\eta,kR) / Rg_l(R)$ is independent of $R$ is that $\log_{10} |R g_l(R)|^2$ is proportional to $\sqrt{ \mathcal{A} (A_c^{1/3}+A_d^{1/3}) Z_c Z_d }$, i.e., $\eta x$ \cite{UDL1, UDL2}. This implication also depends on the linearity assumption and that higher $x$ order terms vary smoothly such that they can be treated as a constant \cite{UDL1, UDL2}. However, as we have seen, the linearity assumption is violated and the higher-order terms cannot be treated as a constant for heavy cluster decay modes in $0.6<x<1$, and thus such conclusion is unwarranted.

\section{conclusions} \label{conclusions}

The possibility of a universal Geiger-Nuttall law was born from the belief that $x$ is always small---as first claimed by Gamow and endorsed subsequently by other authors. On such account, all nuclei (light, heavy, proton rich, or neutron rich) and decay modes (light or heavy) would be unified with a single straight line approximating $\sigma(x)$ near $x=0$. The present findings of the variation of $x$ challenge this belief and leave us with a completely different understanding of the validity of the three classes of Geiger-Nuttall laws. With $x$ varying between 0 and 1 across nuclei and decay modes, no single straight line can achieve the universal Geiger-Nuttall law hoped for. Instead, Geiger-Nuttall laws turn out to be different straight lines with different slopes and intercepts locally approximating $\sigma(x)$ within different $x$ intervals which depend on the decay modes and the nuclei investigated. In light of such a view, there is a fundamental theoretical limitation on the extrapolative power of Geiger-Nuttall laws for nuclei and/or decay modes the $x$ of which lies outside the range of original fit since the approximated straight line would strongly deviate from $\sigma(x)$. This point should be taken into account by future studies that aim to use Geiger-Nuttall laws to systematically study half-life across the nuclear chart and different decay modes. The simple Geiger-Nuttall law and its generalizations (e.g., Viola-Seaborg law) are valid not because $x \approx 0$ but rather because $x$ varies within the interval $x \le 0.6$ where $\sigma(x)$ can be linearly approximated with reasonable accuracy. The extrapolations across the nuclear chart using WS$4^{\text{RBF}}$ show that $x < 0.6$ for almost all nuclei and thus we expect the linearity assumption to hold and the simple Geiger-Nuttall law or its generalizations (i.e., encompassing more than one isotopic chain) to provide an accurate description of $\alpha$ decay of nuclei yet to be observed in the future. The validity of the generalized Geiger-Nuttall laws (e.g., UDL, and UF) encompassing various decay modes can be understood on similar grounds although the present paper shows they are not as universal or general as previously thought. Most significantly, in the regime of the cluster decay $0.6<x<1$, $\sigma(x)$ starts to become noticeably non-linear (especially at $x>0.8$) such that no generalized Geiger-Nuttall law can describe heavy cluster decay. In the ongoing attempts in the unification of decay modes and/or nuclei and description of heavy cluster decay, the present findings indicate that it might be necessary to go beyond the Geiger-Nuttall $Q^{-1/2}$ dependence and incorporate additional higher-order terms proportional to $Q^{1/2}$ (i.e., the $\eta x^2$ term) and/or $Q$ (i.e., the $\eta x^3$ term) to capture the non-linearity of $\sigma(x)$.

\begin{acknowledgments}
The authors would like to thank the American University in Cairo for supporting this research.
\end{acknowledgments}



\FloatBarrier

\appendix* \label{assumptions}

 \section{ \uppercase{Assumptions about} $v$ \uppercase{and} $\zeta_l$ } \label{assumptions}

 \begin{figure}[htp!]
 \centering

  \includegraphics[width= 0.5\textwidth,keepaspectratio]{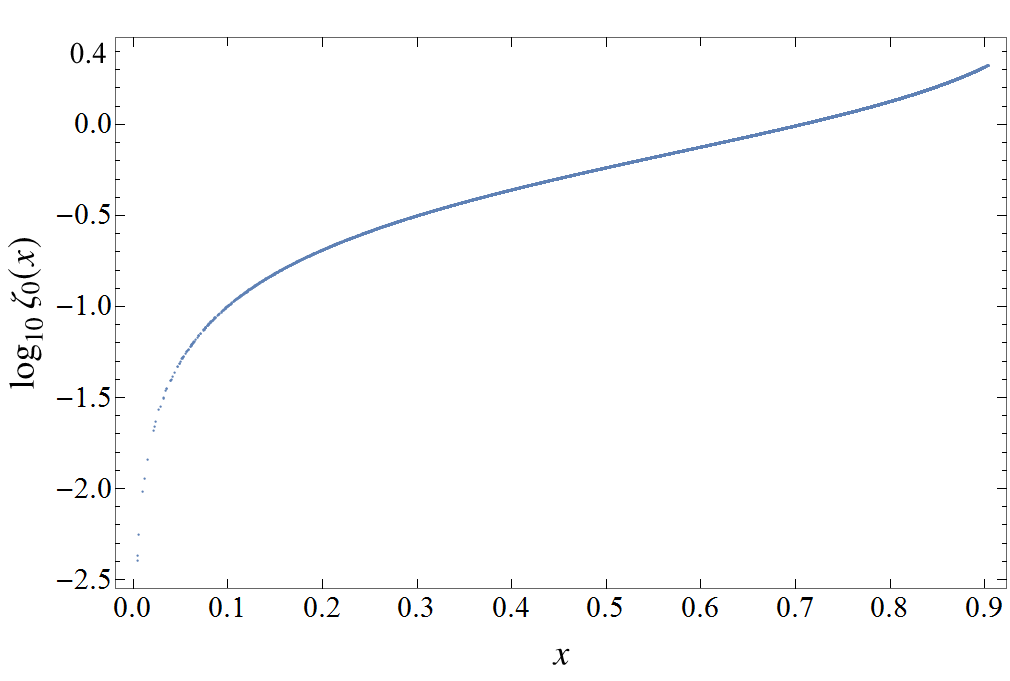}%

      \caption{ $\log_{10}\zeta_0$ vs $x$ for cluster-daughter systems with 12 clusters from the proton up to $\isotope[54]{Ti}$ and nuclei between $78 \le Z \le 132$ and $162 \le A \le 350$.}
        \label{zeta}

\end{figure} 

Here, we show that the quantity $d=\log_{10}  \zeta_l(x) \ln2 /v $ in Eq. \eqref{logT} is constant. By substituting the expression for $|H_{l}^{+}(\eta, k R)|^2$ [Eq. \eqref{CH}] in $T_{1/2}$ [Eq. \eqref{T}] we get

\begin{equation} \label{Tsub}
T_{1/2}=\dfrac{ \zeta_l(x) \exp[2 \gamma_l(x)] \ln 2}{v |R g_l(R)|^2}  
\end{equation} 

It is well known that the most important quantity that determines $T_{1/2}$ is $\exp[2 \gamma_l(x)]$ compared to others. This is because 1) $\exp[2 \gamma_l(x)]$ is very large in the range of tens of orders of magnitude for the experimental region \cite{raredecay} and 2) it varies greatly by tens of orders of magnitude within a single decay mode for a small set of neighboring nuclei (e.g., for $\alpha$ decay of isotopes between $\isotope{Pb}$ and $\isotope{U}$, it ranges between $10^{39}$ and $10^{14}$) \cite{preform1}.

$v$, on the other hand, varies smoothly and retains roughly the same order of magnitude of about $10^{22} \si{fm/s}$ for all decay modes and nuclei so as to be considered a constant. This can be seen from the definition of $v$:

\begin{equation} \label{v}
v=\sqrt{\dfrac{2Q}{\mu}}= 1.38 \times 10^{22} \sqrt{ \dfrac{Q}{\mathcal{A} } }  \si{ \femto \meter / \second}
\end{equation} 

where we have used $\mu= m \mathcal{A}$ where $m= 938.9/c^2$ MeV is the nucleon mass and $c=2.9979 \times 10^{23} \si{ \femto \meter / \second}$ is the speed of light and $\mathcal{A}=A_c A_d/(A_c +A_d)$ is the dimensionless reduced mass. The ratio $ \sqrt{ Q / \mathcal{A} }$ varies smoothly for all decay modes where both $Q$ and $\mathcal{A}$ increase for larger decay modes such that their ratio and hence $v$ remain essentially constant. For instance,  in $\alpha$ decay for nuclei with $78 \le Z \le 132$, $\mathcal{A}\approx4$ for all nuclei and $Q$ is between 1 and 15 MeV such that $ \sqrt{ Q / \mathcal{A} }$ varies between 0.5 and 2. Our calculations for the 12 clusters from the proton up to $\isotope[54]{Ti}$ confirm this. $Q$ is calculated as described in Sec. \ref{results}. We found that for all clusters and nuclei, $v$ is in the order of $10^{22}  \si{ \femto \meter / \second}$. Therefore, we conclude that $v$ is constant as assumed in Sec. \ref{model}. This is consistent with other studies which show that the assault frequency $\nu_a=v/2R$ is roughly a constant equal to $10^{21} \mathrm{s}^{-1}$ for the whole range of nuclei and decay modes \cite{nu1,nu2,nu3,nu4,preform1}. 

$\zeta_l(x)$ [Eq. \eqref{zeta}] is small compared to $\exp[2 \gamma_l(x)]$ and varies smoothly for all decay modes across the nuclear chart. In Fig. \ref{zeta} we plotted $ \log_{10}\zeta_0(x)$ (where we set $l=0$ since it is unavailable for most nuclei considered) vs $x$ for the all cluster-daughter systems. $\log_{10}\zeta_0(x)$ varies very weakly vs $x$ with virtually all cluster-daughter systems lying between 0.5 and -1. The second term containing the $l$ dependence in $\zeta_l(x)$ [Eq. \eqref{zeta}] is small compared to the first term and hence $\zeta_0(x) \approx \zeta_l(x)$. Therefore, $\log_{10} \zeta_l(x)$ is roughly a constant or more precisely, an approximately linear function (for $0.1 \le x \le 0.95$) that depends weakly on $x$.

\FloatBarrier

\input{GNdraft.bbl}
\end{document}
%

%% file: GNdraft.bbl
\providecommand{\noopsort}[1]{}\providecommand{\singleletter}[1]{#1}%